\documentstyle[preprint,aps,epsf]{revtex}	
\begin{document}
%
%
\preprint{$
\begin{array}{l}
\mbox{BA-00-08}\\[-3mm]
\mbox{FERMILAB-Pub-00/037-T}\\[-3mm]
\mbox{hep-ph/0002155}\\[-3mm]
\mbox{January 2000} \\   [2.mm]
\end{array}
$}
\title{Explicit SO(10) Supersymmetric Grand Unified Model\\ 
	for the Higgs and Yukawa Sectors}
\author{Carl H. Albright$^1$ and S.M. Barr$^2$}
\address{$^1$Department of Physics, Northern Illinois University, DeKalb, 
        IL 60115\\
        and \\
        Fermi National Accelerator Laboratory, P.O. Box 500, Batavia, IL 
        60510\\
	$^2$Bartol Research Institute, University of Delaware,
        Newark, DE 19716}
\maketitle
\begin{abstract}

A complete set of fermion and Higgs superfields is introduced with 
well-defined $SO(10)$ properties and $U(1) \times Z_2 \times Z_2$ family 
charges from which the Higgs and Yukawa superpotentials are constructed.  
The structures derived for the four Dirac fermion and right-handed Majorana 
neutrino mass matrices coincide with those previously obtained from
an effective operator approach.  Ten mass matrix input parameters 
accurately yield the twenty masses and mixings of the quarks and leptons 
with the bimaximal atmospheric and solar neutrino vacuum solutions
favored in this simplest version.
\\[0.1in]
PACS numbers: 12.15Ff, 12.10.Dm, 12.60.Jv, 14.60.Pq\\
\end{abstract}
%
%
%
\indent In a series of recent papers \cite{ab1} - \cite{ab5} the authors 
have shown 
how fermion mass matrices can be constructed in an $SO(10)$ supersymmetric
grand unified framework by use of a minimal Higgs structure which solves the 
doublet-triplet splitting problem \cite{b-r}.  The construction was carried 
out in an effective operator approach with phenomenological input, including
the Georgi-Jarlskog relations \cite{g-j}.  Here we show how one can introduce 
a set of matter and Higgs $SO(10)$ superfields with $U(1) \times 
Z_2 \times Z_2$ family charges from which the derived Higgs and Yukawa 
superpotentials uniquely give the structure of the fermion mass matrices 
previously obtained.  The quark and lepton mass and mixing data are 
reproduced remarkably well with the solar neutrino vacuum solution 
preferred, provided the up quark mass is not zero at the GUT scale -- 
otherwise the small angle MSW solution \cite{MSW} is obtained.  The 
right-handed Majorana neutrino matrix arises from a Higgs field which 
couples pairs of superheavy conjugate neutrino singlets.

We begin with a listing in Table I. of the Higgs and matter superfields in the 
proposed model along with their\\
$$\begin{tabular}{ll}
\multicolumn{2}{l}{\bf Higgs\ Fields\ Needed\ to\ Solve\ 
	the\ 2-3\ Problem:}\\[0.1in]
  	${\bf 45}_{B-L}$: & $A(0)^{+-}$ \\
  	${\bf 16}$: & $C(\frac{3}{2})^{-+},\ 
		C'(\frac{3}{2}-p)^{++}$\\
  	$\overline{\bf 16}$: & $\bar{C}(-\frac{3}{2})^{++},
		\ \bar{C}'(-\frac{3}{2}-p)^{-+}$\\
 	${\bf 10}$: & $T_1(1)^{++},\ T_2(-1)^{+-}$\\
  	${\bf 1}$: & $X(0)^{++},\ P(p)^{+-},\ Z_1(p)^{++},
		\ Z_2(p)^{++}$\\[0.1in]
\multicolumn{2}{l}{\bf Additional Higgs\ Fields\ for\ the\ Mass\ Matrices:}
	\\[0.1in]
  	${\bf 10}$: & $T_0(1+p)^{+-},\ T'_o(1+2p)^{+-}$,\\
		& \ $\bar{T}_o(-3+p)^{-+}, \bar{T}'_o(-1-3p)^{-+}$\\
  	${\bf 1}$: & $Y(2)^{-+},\ Y'(2)^{++},\ 
		S(2-2p)^{--},\ S'(2-3p)^{--}$,\\
		& $V_M(4+2p)^{++}$\\[0.1in]
\multicolumn{2}{c}{Table I.  Higgs superfields in the proposed model.}
\end{tabular}$$
\noindent
family charges.  As demonstrated in \cite{b-r}, in order to do all
the symmetry breaking, one ${\bf 45}$ adjoint Higgs with its VEV pointing 
in the $B-L$ direction, a pair of 
${\bf 16} + {\overline{\bf 16}}$ spinor Higgs, plus a pair
of ${\bf 10}$ vector Higgs and several Higgs singlets are required.  In 
order to complete the 
construction of the Dirac mass matrices, four more vector Higgs
and four additional singlets are needed.  Finally, one
Higgs singlet is introduced to generate the right-handed Majorana mass 
matrix.  

From the Higgs $SO(10)$ and family assignments, it is then possible to 
write down explicitly the full Higgs superpotential, where we have written
it as the sum of five terms:

\begin{equation}
$$\begin{array}{rcl} 
        W_{\rm Higgs} &=& W_A + W_{CA} + W_{2/3} + W_{H_D} + W_R\\[0.1in]
	W_A &=& tr A^4/M + M_A tr A^2\\[0.1in]
	W_{CA} &=& X(\overline{C}C)^2/M^2_C + F(X) \\
	 & & + \overline{C}'(PA/M_1 + Z_1)C + \overline{C}(PA/M_2 + 
	  Z_2)C'\\[0.1in]
	W_{2/3} &=& T_1 A T_2 + Y' T^2_2\\[0.1in]
	W_{H_D} &=& T_1 \overline{C}\overline{C} Y'/M + 
	  \overline{T}_0 C C' + \overline{T}_0(T_0 S + T'_0 S')\\[0.1in]
	W_R &=& \overline{T}_0 \overline{T}'_0 V_M\\
\end{array}$$
\end{equation}

\noindent
The Higgs singlets are all assumed to develop VEV's at the GUT scale.
$W_A$ fixes $\langle A \rangle$ through the $F_A = 0$ condition where one 
solution is $\langle A \rangle \propto B-L$, the Dimopoulos-Wilczek solution
\cite{d-w}.  $W_{CA}$ gives a GUT-scale VEV to $\overline{C}$ and $C$ by 
the $F_X = 0$ condition and also couples the adjoint $A$ to the spinors 
$C,\ \overline{C},\ C'$ and $\overline{C}'$ without destabilizing the 
Dimopoulos-Wilczek solution or giving Goldstone modes.  $W_{2/3}$ gives the 
doublet-triplet splitting by the Dimopoulos-Wilczek mechanism.  $W_{H_D}$
mixes the $(1,2,-1/2)$ doublet in $T_1$ with those in $C'$ 
(by $F_{\overline{C}} = 0$), and in $T_0$ and $T'_0$ (by $F_{\overline{T}_0} 
= 0$).
To fill out the model, we specify the $SO(10) \times U(1) \times Z_2 
\times Z_2$ quantum numbers of the various matter fields in Table II.
We require three chiral spinor fields ${\bf 16}_i$, one for each light family,
two vector-like pairs of ${\bf 16} + \overline{\bf 16}$ spinors which 
can get superheavy, a pair of superheavy ${\bf 10}$ fields in the vector 
representation, and three pairs of superheavy ${\bf 1} - {\bf 1}^c$ fermion 
singlets.  

$$\begin{tabular}{lll}
	${\bf 16}_1(-\frac{1}{2}-2p)^{+-}$ \  & 
		${\bf 16}_2(-\frac{1}{2}+p)^{++}$ \ & 
		${\bf 16}_3(-\frac{1}{2})^{++}$ \\
	${\bf 16}(-\frac{1}{2}-p)^{-+}$ \ & 
		${\bf 16}'(-\frac{1}{2})^{-+}$ \\
	$\overline{\bf 16}(\frac{1}{2})^{+-}$ \ &
	$\overline{\bf 16}'(-\frac{3}{2}+2p)^{+-}$\\[0.1in]
	${\bf 10}_1(-1-p)^{-+}$ \ & ${\bf 10}_2(-1+p)^{++}$
	  \\[0.1in]
	${\bf 1}_1(2+2p)^{+-}$ \ & 
		${\bf 1}_2(2-p)^{++}$ \  
		& ${\bf 1}_3(2)^{++}$\\[0.1in]
	${\bf 1}^c_1(-2-2p)^{+-}$ \ & ${\bf 1}^c_2(-2)^{+-}$
		\ & ${\bf 1}^c_3(-2-p)^{++}$ \\[0.1in]
\multicolumn{3}{c}{Table II.  Matter superfields in the proposed model.}
  \end{tabular}$$

In terms of these fermion fields and the Higgs fields previously introduced,
one can then spell out all the terms in the Yukawa superpotential which 
follow from their $SO(10)$ and $U(1) \times Z_2 \times Z_2$ assignments:

\begin{equation}
$$\begin{array}{rl}
        W_{Yukawa} &= {\bf 16}_3 \cdot {\bf 16}_3 \cdot T_1 + {\bf 16}_2
                \cdot {\bf 16} \cdot T_1
                + {\bf 16}' \cdot {\bf 16}' \cdot T_1\\
                &+ {\bf 16}_3 \cdot {\bf 16}_1 \cdot T'_0
                + {\bf 16}_2 \cdot {\bf 16}_1 \cdot T_0
                + {\bf 16}_3 \cdot {\overline{\bf 16}} \cdot A\\
                &+ {\bf 16}_1 \cdot \overline{\bf 16}' \cdot Y'
                + {\bf 16} \cdot {\overline{\bf 16}} \cdot P
                + {\bf 16}' \cdot {\overline{\bf 16}}' \cdot S\\
                &+ {\bf 16}_3 \cdot {\bf 10}_2 \cdot C'
                + {\bf 16}_2 \cdot {\bf 10}_1 \cdot C
                + {\bf 10}_1 \cdot {\bf 10}_2 \cdot Y\\
                &+ {\bf 16}_3 \cdot {\bf 1}_3 \cdot \overline{C}
                + {\bf 16}_2 \cdot {\bf 1}_2 \cdot \overline{C}
                + {\bf 16}_1 \cdot {\bf 1}_1 \cdot \overline{C}\\
                &+ {\bf 1}_3 \cdot {\bf 1}^c_3 \cdot Z
                + {\bf 1}_2 \cdot {\bf 1}^c_2 \cdot P
                + {\bf 1}_1 \cdot {\bf 1}^c_1 \cdot X\\
                &+ {\bf 1}^c_3 \cdot {\bf 1}^c_3 \cdot V_M
                + {\bf 1}^c_1 \cdot {\bf 1}^c_2 \cdot V_M\\
  \end{array}$$
\end{equation}

\noindent 
where the coupling parameters have been suppressed.  To obtain the GUT
scale structure for the fermion mass matrix elements, all but the three
chiral spinor fields in the first line of Table II. will be integrated out.  
The right-handed Majorana 
matrix elements will all be generated through the Majorana couplings of 
the $V_M$ Higgs field with conjugate singlet fermions as given above.

With R-parity conserved, $d = 4$ proton decay operators are forbidden.
The  $d = 5$ proton decay operators induced by colored-Higgsino exchange
that are generally present in unified models are present here but are not
dangerous.  It can be shown that the family charge assignments prevent
any new and dangerous proton decay operators from arising.

The procedure for deriving the Dirac mass matrices $U,\ D,\ L$, and $N$
is the following.  For each type of fermion $f$, where $f = u_L,\ u^c_L,
\ d_L,\ d^c_L,\ \ell^-_L,\ \ell^+_L,\ \nu_L$ and $\nu^c_L$, the superheavy
mass matrix connecting the $f$ to the $SU(3) \times SU(2) 
\times U(1)$-conjugate representation $\overline{f}$ is first found from
Eq. (2) by setting the weak-scale VEV's and the intermediate-scale VEV,
$V_M$, to zero.  This will give three zero mass eigenstates for each type 
of $f$, corresponding to the three light families.  
Then the terms in Eq. (2) involving $\langle T_1 \rangle,\ \langle C' 
\rangle,\ \langle T_0 \rangle$, and $\langle T'_0 \rangle$ give rise to the 
$3 \times 3$ Dirac mass matrices coupling $u_L$ to $u^c_L$, etc.
This procedure is spelled out explicitly in \cite{ab6}.

Under the assumption that the zero-mass states have their large components
in the chiral representations ${\bf 16}_1,\ {\bf 16}_2$ and ${\bf 16}_3$, 
and all the other components are small, the Dirac mass matrices obtained have
precisely the structure previously found in our studies by means
of an effective operator approach:

\begin{equation}
\begin{array}{ll}
U = \left( \begin{array}{ccc} \eta & 0 & 0 \\
  0 & 0 & \epsilon/3 \\ 0 & - \epsilon/3 & 1 \end{array} \right), 
  & D = \left( \begin{array}{ccc} 0 & \delta & \delta' e^{i\phi}
  \\
  \delta & 0 & \sigma + \epsilon/3  \\
  \delta' e^{i \phi} & - \epsilon/3 & 1 \end{array} \right), \\ & \\
N = \left( \begin{array}{ccc} \eta & 0 & 0 \\ 0 & 0 & - \epsilon \\
  0 & \epsilon & 1 \end{array} \right),
  & L = \left( \begin{array}{ccc} 0 & \delta & \delta' e^{i \phi} \\
  \delta & 0 & -\epsilon \\ \delta' e^{i\phi} & 
  \sigma + \epsilon & 1 \end{array} \right),
\end{array}
\end{equation}

\noindent
with $U$ and $N$ in units of $M_U$ and $D$ and $L$ in units of $M_D$. 
The matrix parameters are identified with the Yukawa couplings and 
Higgs couplings and VEV's as follows:

\begin{equation}
$$\begin{array}{rlrl}
  M_U =& (t_3)_{5(10)}, & M_D =& (t_3)_{\bar{5}(10)},\\
  \epsilon M_U =& |3(a_q/p)(t_2)_{5(10)}|, & \epsilon M_D =& 
	|3(a_q/p)(t_2)_{\bar{5}(10)}|,\\
  \eta M_U =& (y'/s'')^2 (t')_{5(10)}, & \sigma M_D =& 
	-(c/y)(c')_{\bar{5}(16)},\\
  	& & \delta M_D =& t_0 \bar{t}_0 /s,\\
	& & \delta' M_D =& (t'_0 \bar{t}_0 /s')e^{-i\phi},\\
  \end{array}$$
\end{equation}

\noindent
where the subscripts on $t_2,\ t_3,\ t'$ and $c'$ refer to the 
SU(5)[SO(10)] representation content of the VEV's. The following 
shorthand notation has been introduced

\begin{equation}
$$\begin{array}{rlrl}
  t_3 =& \lambda_{16_3 16_3 T_1}\langle T_1 \rangle, 
	&\quad t_2 =& \lambda_{16_2 16 T_1}\langle T_1 \rangle,\\ 
  t' =& \lambda_{16' 16' T_1} \langle T_1 \rangle, &\quad c' =& 
	\lambda_{16_310_2C'}\langle C' \rangle,\\
  c =& \lambda_{16_210_1C}\langle C \rangle, &\quad \bar{c}_i =& 
        \lambda_{16_i1_i\bar{C}}\langle \bar{C} \rangle, 
        i=1,2,3,\\
  p =& \lambda_{16 \overline{16} P}\langle P \rangle, &\quad p_{22} =& 
	\lambda_{1_2 1^c_2 P}\langle P \rangle,\\
  a_q =& \lambda_{16_3 \overline{16} A}\langle A \rangle_{B = 1/3}, 
	&\quad x =& \lambda_{1_1 1^c_1 X}\langle X \rangle,\\
  y =& \lambda_{10_1 10_2 Y}\langle Y \rangle, &\quad y' =& 
	\lambda_{16_1 \overline{16}' Y'}\langle Y' \rangle,\\
  z =& \lambda_{1_3 1^c_3 Z}\langle Z \rangle, &\quad s =& 
	\lambda_{T_0 \bar{T}_0 S}\langle S \rangle,\\
  s' =& \lambda_{T'_0 \bar{T}_0 S'}\langle S' \rangle, &\quad s'' =&  
	\lambda_{16' \overline{16}' S}\langle S \rangle,\\
  t_0 =& \lambda_{16_116_2T_0}, &\quad t'_0 =& \lambda_{16_1 16_3T'_0}\\
  \bar{t}_0 =& \lambda_{CC'\bar{T}_0}\langle C \rangle \langle C' \rangle.\\
  \end{array}$$
\end{equation}

The parameter $\eta$ is introduced to give a tiny non-zero mass to the up 
quark at the $\Lambda_G$ scale.  Its appearance in $N$ will also play an 
important 
role in the determination of the type of solar neutrino solution.  It should
also appear in $D$ and $L$ but its effect is negligibly small there and of 
no consequence, so it is dropped.  The only phase then appearing in the 
matrices is $\phi$ associated with $\delta'$, as other phases are unphysical
and can be rotated away with the exception of that associated with $\epsilon$.
It turns out, however, that the best fits to the data prefer a real $\epsilon$.
Hence $\phi$ which can be identified with the complexity of the VEV of the 
$S'$ Higgs singlet is solely responsible for CP-violation in the quark sector.
The structures of the matrix elements given in Eqs. (3), (4) and (5) can 
be understood in terms of simple Froggatt-Nielsen diagrams \cite{f-n} 
given in \cite{ab6}.

Note that the 33 elements of the Dirac mass matrices are scaled by the VEV's
of the ${\bf 10},\ T_1$.  But the $F = 0$ conditions for the Higgs 
superpotential require that the pair of Higgs doublets which remain light down
to the electroweak scale arise from $5(T_1),\ \bar{5}(T_1),\ \bar{5}(C')$
and, to a very small extent from $T_0$ and $T'_0$ terms, which are ignored 
here.  In particular, we can write in terms of a mixing angle $\gamma$

\begin{equation}
	H_U = 5(T_1),\quad H_D = \bar{5}(T_1)\cos \gamma + 
		\bar{5}(C')\sin \gamma\\
\end{equation}

\noindent
whereas the orthogonal combination has become superheavy at the GUT scale.
Thus the ratio of the 33 mass matrix elements found from Eqs. (4) and (6) is 
given in terms of the VEV's, $v_u$ and $v_d$ of $H_U$ and $H_D$, respectively,
by 

\begin{equation}
	M_U/M_D = v_u/(v_d \cos \gamma) \equiv \tan \beta /\cos \gamma\\
\end{equation}

\noindent
Hence we find that the large $M_U/M_D$ ratio required for the top to 
bottom quark masses can be achieved with a {\it moderate} $\tan \beta$ 
provided $\cos \gamma$ is small.

Turning to the right-handed Majorana mass matrix, we use the zero mass 
left-handed conjugate states that were found implicitly above for the
Dirac matrix $N$ to form the basis for $M_R$.  
The right-handed Majorana neutrino matrix is then obtained from the last
two terms in Eq. (2), and we find 

\begin{equation}
$$ M_R = \left(\matrix{ 0 & A\epsilon^3 & 0 \cr A\epsilon^3 & 0 & 0 \cr 
	0 & 0 & 1 \cr}\right)\Lambda_R$$
\end{equation}

\noindent 
where 

\begin{equation}
$$\begin{array}{rl}
  \Lambda_R =& \lambda_{1^c_3 1^c_3 V_M} \langle V_M \rangle (\bar{c}_3/z)^2,\\
  A\epsilon^3\Lambda_R =& \lambda_{1^c_1 1^c_2 V_M} \langle V_M \rangle 
	(\bar{c}_1/x)(\bar{c}_2/p_{22})\\
  \end{array}$$
\end{equation}

\noindent 
Note that the whole right-handed Majorana mass matrix has been generated 
in this simple model by the one Majorana VEV coupling superheavy conjugate 
fermion singlets.  By means of the seesaw formula \cite{gm-r-s}, one can 
then compute the light neutrino mass matrix 

\begin{equation}
$$M_\nu = N^T M_R^{-1} N = \left(\matrix{ 0 & 0 & -\frac{\eta}{A\epsilon^2}\cr
	0 & \epsilon^2 & \epsilon\cr 
	-\frac{\eta}{A\epsilon^2} & \epsilon & 1\cr}\right)
	M^2_U/\Lambda_R$$
\end{equation}

We now address the predictions of the mass matrices.  For this purpose
it is convenient to replace the parameters $\delta$ and $\delta'$ by

\begin{equation}
	t_Le^{i\theta} \equiv {{\delta - \sigma\delta'e^{i\phi}}
		\over{\sigma\epsilon/3}},\quad 
	t_R \equiv {{\delta\sqrt{\sigma^2 + 1}}\over{\sigma\epsilon/3}}\\
\end{equation}

\noindent 
which are essentially the left-handed and right-handed Cabibbo angles.
In terms of the dimensionless parameters $\epsilon,\ \sigma,\ t_L,
\ t_R,\ e^{i\theta},\ \eta,\ A$, and $M_U/M_D$, we then find at the GUT scale

\begin{equation}
\begin{array}{rl}
m^0_t/m^0_b \cong & (\sigma^2 + 1)^{-1/2}M_U/M_D,\quad
	m_u^0/m_t^0 \cong \eta, \\
m_c^0/m_t^0 \cong & \frac{1}{9} \epsilon^2 \cdot [1 - \frac{2}{9} 
	\epsilon^2],\quad 
	m_b^0/m_{\tau}^0 \cong 1 - \frac{2}{3} \frac{\sigma}{\sigma^2 + 1} 
	\epsilon, \\
m_s^0/m_b^0 \cong & \frac{1}{3} \epsilon 
	\frac{\sigma}{\sigma^2 + 1}\cdot [1  + \frac{1}{3} \epsilon 
	\frac{1 - \sigma^2 - \sigma \epsilon/3}
	{\sigma (\sigma^2 + 1)} + \frac{1}{2} (t_L^2 + t_R^2) ], \\
m_d^0/m_s^0 \cong & t_L t_R \cdot [ 1 - \frac{1}{3} 
	\epsilon\frac{\sigma^2 + 2}{\sigma (\sigma^2 + 1)} - (t_L^2 + t_R^2)\\
	&\quad + (t_L^4 + t_L^2 t_R^2 + t_R^4)], \\
m_{\mu}^0/m_{\tau}^0 \cong & \epsilon 
	\frac{\sigma}{\sigma^2 + 1} \cdot[1 + \epsilon \frac{1 - 
	\sigma^2 - \sigma \epsilon}{\sigma(\sigma^2 + 1)}
	+ \frac{1}{18}(t_L^2 + t_R^2)], \\
m_e^0/m_{\mu}^0 \cong & \frac{1}{9} t_L t_R \cdot 
	[1 - \epsilon \frac{\sigma^2 + 2}{\sigma(\sigma^2 + 1)} + \epsilon^2
	\frac{\sigma^4 + 9 \sigma^2/2 + 3}{\sigma^2(\sigma^2 + 1)^2}\\
	& \quad - \frac{1}{9} (t_L^2 + t_R^2)], \\
V_{cb}^0 \cong & \frac{1}{3} \epsilon 
	\frac{\sigma^2}{\sigma^2 + 1} \cdot [1 + \frac{2}{3} 
	\epsilon \frac{1}{\sigma(\sigma^2 + 1)}], \\
V_{us}^0 \cong & t_L [ 1 -\frac{1}{2} t_L^2 - t_R^2 
	+ t_R^4 + \frac{5}{2} t_L^2 t_R^2 + \frac{3}{8} t_L^4\\ 
	& \quad - \frac{\epsilon}{3 \sigma \sqrt{\sigma^2 + 1}} \frac{t_R}{t_L} 
	e^{- i \theta}], \\
V_{ub}^0 \cong & \frac{1}{3} t_L \epsilon 
	\frac{1}{\sigma^2 + 1} [\sqrt{\sigma^2 + 1} \frac{t_R}{t_L} 
	e^{-i \theta} (1 - \frac{1}{3} \epsilon \frac{\sigma}{\sigma^2 + 1})\\
	& \quad - (1 - \frac{2}{3}\epsilon \frac{\sigma}{\sigma^2 + 1})], \\
m_2^0/m_3^0 \cong & \left(\frac{\eta}{A\epsilon\sqrt{1 + \epsilon^2}}\right)
	\left[ 1 + \frac{\eta}{A\epsilon^3\sqrt{1 + \epsilon^2}} \right], \\ 
m_1^0/m_3^0 \cong & \left(\frac{\eta}{A\epsilon\sqrt{1+\epsilon^2}}
	\right)\left[ 1 - \frac{\eta}{2A\epsilon^3\sqrt{1+\epsilon^2}} 
	\right],\\
U_{\mu 3}^0 \cong & - \frac{1}{\sqrt{\sigma^2 + 1}}(\sigma - \epsilon
	\frac{\sigma^2}{\sigma^2 + 1}),\\[0.05in]
U_{e 2}^0 \cong & -\frac{1}{\sqrt{2}}\left[ 1 - \frac{\epsilon}{3\sigma}t_L 
	e^{i\theta}\right.\\
	& \quad + \left. \frac{1}{3\sqrt{\sigma^2 + 1}}(1 + \epsilon\sigma)
	t_R\right],\\
U_{e 3}^0 \cong & \frac{1}{3\sqrt{\sigma^2 + 1}}(\sigma - \epsilon)
	t_R - \frac{\eta}{A\epsilon^2}\\
\end{array}
\end{equation}

\noindent
Note that the Georgi-Jarlskog relations \cite{g-j}, $m^0_s \cong \frac{1}{3}
m^0_\mu$ and $m^0_d \cong 3m^0_e$, emerge as required by design.  
The quark and charged lepton data are best fit at the low scale (see below) 
by assigning the 
following values to the model parameters: $\epsilon = 0.145,\ \sigma = 1.78,
\ t_L = 0.236,\ t_R = 0.205,\ \theta = 34^o\ ({\rm corresponding\ to\ } 
\delta =0.0086,\ \delta' = 0.0079,\ \phi =54^o),\ \eta = 8 \times 10^{-6}$,
and $M_U/M_D \simeq 113$. 

As noted earlier, in order to obtain the simple mass matrices in Eq. (3), 
we had to assume that the zero-mass states have their large components in
the chiral representations ${\bf 16}_1,\ {\bf 16}_2$, and ${\bf 16}_3$.
The conditions on the state normalization factors are all satisfied provided
the following ratios are much less than unity:

\begin{equation}
\begin{array}{rl}
  (a/p)^2,& (y'/s'')^2,\ (c/y)^2,\\[0.1in]
	  &\ (\bar{c}_1/x)^2,\ (\bar{c}_2/p_{22})^2,\ (\bar{c}_3/z)^2 \ll 1\\
\end{array}
\end{equation} 

\noindent
With the numerical choice of parameters given above and near equality of the
various Higgs couplings, we find $(a/p)^2 \simeq 0.02$ and $(y'/s'')^2
\sim 6 \times 10^{-6}$, so the first two conditions are easily satisfied.
Requiring that $(c/y)^2 \ll 1$ and with the expression for $\sigma$ obtained
from Eqs. (4), we find

\begin{equation}
\begin{array}{rl}
\tan \gamma \equiv& {{\langle \bar{5}(C') \rangle}\over{\langle 
	\bar{5}(T_1) \rangle}} \gg \sigma \\[0.1in]
 \tan \beta \simeq& \sqrt{\sigma^2 + 1}(\cos \gamma)m^0_t/m^0_b 
	\ll m^0_t/m^0_b\\
  \end{array}
\end{equation}

\noindent
in terms of the $T_1 - C'$ mixing angle, $\gamma$, in Eq. (6).
With $c/y \cong 0.1$, for example, $\tan \gamma \simeq 18$ which implies
$\tan \beta \simeq 6$, a very reasonable mid-range value allowed 
by experiment.  The others can also be satisfied \cite{ab6}.

In \cite{ab6} we have evolved the results in Eqs. (12) down to the low 
scales with a value for $\tan \beta = 5,\ \Lambda_G = 2 \times 10^{16}$ GeV,
$\Lambda_{SUSY} = m_t(m_t),\ \alpha_s(M_Z) = 0.118,\ \alpha(M_Z) = 1/127.9$,
and $\sin^2 \theta_W = 0.2315$.  With the quantities $m_t(m_t) = 165\ {\rm
GeV},\ m_{\tau} = 1.777\ {\rm GeV},\ m_{\mu} = 105.7\ {\rm MeV},\ m_e = 
0.511\ {\rm MeV},\ m_u = 4.5\ {\rm MeV},\ V_{us} = 0.220,\ V_{cb} = 0.0395$,
and $\delta_{CP} = 64^o$ used to determine the input parameters, 
$M_U \simeq 113\ {\rm GeV},\ M_D \simeq 1\ {\rm GeV}$, and $\sigma,\ \epsilon,
\ t_L, \ t_R,\ \theta$ and $\eta$ given earlier, the following values are 
obtained compared with experiment \cite{data} in parentheses:

\begin{equation}
$$\begin{array}{rll}
        m_c(m_c) =& 1.23\ {\rm GeV}\qquad & (1.27 \pm 0.1\ {\rm GeV})\\
        m_b(m_b) =& 4.25\ {\rm GeV}\qquad & (4.26 \pm 0.11\ {\rm GeV})\\
        m_s({\rm 1\ GeV}) =& 148\ {\rm MeV}\qquad & (175 \pm 50\ {\rm MeV})\\
        m_d({\rm 1\ GeV}) =& 7.9\ {\rm MeV}\qquad & (8.9 \pm 2.6\ {\rm MeV})\\
        |V_{ub}/V_{cb}| =& 0.080 \qquad & (0.090 \pm 0.008)\\
        \end{array}$$
\end{equation}

\noindent
where finite SUSY loop corrections for $m_b$ and $m_s$ have been scaled
to give $m_b(m_b) \simeq 4.25$ GeV for $\tan \beta = 5.$

The effective light neutrino mass matrix of Eq. (10) leads to bimaximal 
mixing with a large angle solution for atmospheric neutrino oscillations 
\cite{atm} and the ``just-so'' vacuum solution \cite{bimax} involving two 
pseudo-Dirac neutrinos, if we set $\Lambda_R = 2.4 \times 10^{14}$ GeV and
$A = 0.05$.  We then find 

\begin{equation}
$$\begin{array}{ll}
  \multicolumn{2}{c}{m_3 = 54.3\ {\rm meV},\ m_2 = 59.6\ {\rm \mu eV},
	\ m_1 = 56.5\ {\rm \mu eV}}\\[0.1in]
  \multicolumn{2}{c}{U_{e2} = 0.733,\ U_{e3} = 0.047,\ U_{\mu 3} = -0.818,
	\ \delta'_{CP} = -0.2^o}\\[0.1in]
        \Delta m^2_{23} = 3.0 \times 10^{-3}\ {\rm eV^2},\quad &        
                \sin^2 2\theta_{atm} = 0.89\\[0.1in]
        \Delta m^2_{12} = 3.6 \times 10^{-10}\ {\rm eV^2},\quad &
                \sin^2 2\theta_{solar} = 0.99\\[0.1in]
        \Delta m^2_{13} = 3.0 \times 10^{-3}\ {\rm eV^2},\quad &
                \sin^2 2\theta_{reac} = 0.009\\
        \end{array}\\$$
\end{equation}

\noindent
The effective scale of the right-handed Majorana
mass contribution occurs two orders of magnitude lower than the SUSY GUT 
scale of $\Lambda_G = 1.2 \times 10^{16}$ GeV.  The effective two-component
reactor mixing angle given above should be observable at a future neutrino 
factory, whereas the present limit from the CHOOZ experiment \cite{CHOOZ} 
is approximately 0.1 for the above $\Delta m^2_{23}$.  In principle, the
parameter $A$ appearing in $M_R$ can also be complex, but we find that in 
no case does the leptonic CP-violating phase, $\delta'_{CP}$ exceed $10^o$
in magnitude.  Hence the model predicts leptonic CP-violation will be 
unobservable.

The vacuum solar solution depends critically on the appearance of the 
parameter $\eta$ in the matrix $N$, corresponding to the non-zero $\eta$ 
entry in $U$ which gives the up quark a mass at the GUT scale.  Should we 
set $\eta = 0$, only the small-angle MSW solution \cite{MSW} would be 
obtained for the solar neutrino oscillations.  The large angle MSW solution 
is disfavored by the larger hierarchy, i.e., very small $A$ value, 
required in $M_R$.  

In summary, we have constructed an explicit $SO(10)$ supersymmetric grand 
unified model for the Higgs and Yukawa superpotentials which reproduces
the fermion mass matrices previously obtained in an effective operator 
approach.  All the quark and lepton mass and mixing data are fit remarkably
well with a $\tan \beta$ in the range of 5 - 10 with matrix parameters 
which are also quite reasonable.

The research of SMB was supported in part by the Department of Energy under 
contract No. DE-FG02-91ER-40626.  Fermilab is operated by Universities 
Research Association Inc. under contract No. DE-AC02-76CH03000 with the 
Department of Energy.\\
%
%

%
\end{document}